\documentclass{jpsj-suppl}
\usepackage{txfonts} 
\usepackage{bm}
\usepackage{bbm}

\def\gtap{\ \raise.3ex\hbox{$>$\kern-.75em\lower1ex\hbox{$\sim$}}\ }
\def\ltap{\ \raise.3ex\hbox{$<$\kern-.75em\lower1ex\hbox{$\sim$}}\ }

\title{Neutrino-induced meson productions}

\author{Satoshi X. \textsc{Nakamura}}

\inst{Department of Physics, Osaka University, Toyonaka, Osaka 560-0043, Japan}

\email{nakamura@kern.phys.sci.osaka-u.ac.jp}

\recdate{October 1, 2015}
\abst{
We develop
a dynamical coupled-channels (DCC) model for neutrino-nucleon
reactions in the resonance region,
by extending the DCC model that we have previously developed through an
analysis of $\pi N, \gamma N\to \pi N, \eta N, K\Lambda, K\Sigma$
reaction data for $W\le 2.1$~GeV.
We analyze electron-induced reaction
data for both proton and neutron targets
to determine the vector current form factors
up to $Q^2\le$ 3.0~(GeV/$c$)$^2$.
Axial-current matrix elements are
derived in accordance with 
the Partially Conserved Axial Current (PCAC)
relation to
the $\pi N$ interactions of the DCC model.
As a result, we can uniquely determine
the interference pattern between resonant
and non-resonant amplitudes.
Our calculated cross sections for neutrino-induced single-pion productions
are compared with available data, and are found to be in reasonable
agreement with the data.
We also calculate the double-pion production cross sections
in the resonance region,
for the first time, with relevant resonance contributions and channel couplings.
The result is compared with the double-pion production data.
For a future development of a
neutrino-nucleus reaction model and/or a neutrino event generator
for analyses of neutrino experiments,
the DCC model presented here can give a useful input.
}

\kword{Neutrino-nucleus interactions, Meson production, Meson-baryon
interactions, neutrino oscillation}

\begin{document}
\maketitle

\section{Introduction}

More improved understanding of neutrino-nucleus reactions is a critical
issue for addressing the leptonic CP violation and the neutrino mass
hierarchy with forthcoming neutrino oscillation experiments.
The neutrino oscillation experiments
utilize neutrinos in a wide energy range, 
and therefore the relevant neutrino-nucleus reactions have various 
microscopic reaction mechanisms depending on the kinematics.
For a relatively low-energy neutrino ($E_\nu\ltap$1~GeV), 
the dominant reaction mechanisms are the quasi-elastic knockout of a
nucleon, and quasi-free excitation of the $\Delta(1232)$
resonance followed by a decay into a $\pi N$ final state. 
For a higher-energy neutrino ($2\ltap E_\nu\ltap$4~GeV),
a large portion of data are from 
higher resonance excitations and deep inelastic scattering (DIS).
In order to understand the neutrino-nucleus reactions of these
different characteristics, obviously, it is important to combine different
expertise.
For example, nuclear theorists and neutrino experimentalists recently organized a
collaboration at the J-PARC branch of the KEK theory
center~\cite{collab,unified} to tackle this challenging problem.

In this work,
we focus on studying the neutrino reactions in the resonance region where
the total hadronic energy $W$ extends, $m_N+m_\pi < W \ltap 2$~GeV; 
$m_N$ ($m_\pi$) is the nucleon (pion) mass.
Furthermore
we will be concerned with the neutrino reaction on a single nucleon. 
In the resonance region, 
particularly between the $\Delta(1232)$ and DIS regions,
we are still in the stage of developing a
single nucleon model that is a basic ingredient to construct a
neutrino-nucleus reaction model. 
Several theoretical models have been developed
for neutrino-nucleon reactions in the resonance region;
particularly
the $\Delta(1232)$ region has been extensively studied because of
its importance.
However, there still remain conceptual and/or practical problems in the existing
models as follows:
First, we point out that 
reactions in the resonance region are
multi-channel processes in nature.
However, no existing model takes account of
the multi-channel couplings required by the unitarity.
Second, the neutrino-induced double pion productions over the entire resonance
     region have not been seriously studied previously,
even though their production rates are expected to be 
comparable or even more important than those for the single-pion productions
around and beyond the second resonance region.
Third,
interference between resonant and non-resonant amplitudes are not 
well under control for the axial-current in most of the
previous models.

Our goal here is to develop a neutrino-nucleon reaction model in the
resonance region by overcoming the problems discussed above.
In order to do so, the best available option would be to work with a
coupled-channels model.
In the last few years, we have developed a dynamical coupled-channels (DCC)
model to analyze $\pi N, \gamma N\to \pi N, \eta N, K\Lambda, K\Sigma$
reaction data for a study of the baryon spectroscopy~\cite{knls13}.
In there, we have shown that 
the model is successful in giving
a reasonable fit to a large amount ($\sim$ 23,000 data points) of the data.
The model also has been shown to give a reasonable prediction for
the pion-induced double pion productions~\cite{kamano-pipin}.
Thus the DCC model seems a promising starting point for developing a
neutrino-reaction model in the resonance region.
At $Q^2=0$, we already have made an extension of the DCC model to the
neutrino sector by invoking
the PCAC (Partially Conserved Axial Current) hypothesis~\cite{knls12}.
At this particular kinematics, the cross section is given by the
divergence of the axial-current amplitude that is related to the $\pi N$ amplitude
through the PCAC relation.
However, for describing the neutrino reactions in the whole kinematical
region ($Q^2\ne 0$), a dynamical model for the vector- and
axial-currents is needed.

Practically,
we need to do the following tasks for extending the DCC model to
cover the neutrino reactions.
Regarding the vector current, we already have fixed the amplitude for the
proton target at $Q^2$=0 in our previous analysis~\cite{knls13}.
The remaining task is to determine the $Q^2$-dependence of the vector
couplings, i.e., form factors.
This can be done by analyzing data for the single pion
electroproduction and inclusive electron scattering.
A similar analysis also needs to be done with the neutron target model.
By combining the vector current amplitudes for the proton and neutron targets, we
can do the isospin separation of the vector current.
This is a necessary step before calculating neutrino processes.
As for the axial-current matrix elements at $Q^2$=0,
we derive them so that the consistency, required by the PCAC relation,
with the DCC $\pi N$ interaction model is maintained.
As a result of this derivation, 
the interference pattern between the resonant and non-resonant
amplitudes are uniquely fixed within our model;
this is an advantage of our approach.
For the $Q^2$-dependence of the axial-current matrix elements,
we still inevitably need to employ a simple ansatz due to
the lack of experimental information.
This is a limitation shared by all the existing neutrino-reaction models
in the resonance region.

With the vector- and axial-currents as described above, 
we calculate cross sections for the neutrino-induced meson productions in the resonance
region.
We compare our numerical results with available data for
single-pion and double-pion productions.
Particularly, comparison with the double-pion production data is
made for the first time with the relevant resonance contributions 
and coupled-channels effects taken into account.
For a fuller presentation of this work, we refer the readers to Ref.~\cite{dcc-neutrino}.

\section{Formalism}
\label{sec:xsc}


The weak interaction Lagrangian for charged-current (CC) processes is given by
\begin{equation}
   {\cal L}^{\rm CC} =
{G_F V_{ud}\over \sqrt{2}}\int d^3x [ J^{\rm CC}_\mu(x) l^{\rm CC\, \mu}(x) + {\rm h.c.} ]
\ ,
\label{eq:weak_lag}
\end{equation}
where $G_F$ is the Fermi coupling constant and 
$V_{ud}$ is the CKM matrix element.
The leptonic current is denoted by $l^{\rm CC}_\mu$,
while the hadronic current is
\begin{eqnarray}
J^{\rm CC}_\mu(x)=V^{+}_\mu(x) - A^{+}_\mu(x) \ ,
\label{eq:J}
\end{eqnarray}
where $V^{+}_\mu$ and $A^{+}_\mu$ are the vector and axial currents.
The superscript $+$ denotes the isospin raising operator.

\subsection{Matrix elements of non-resonant currents}
\label{sec:nonres}

As in Eqs.~(\ref{eq:J}), the current consists of the
vector and axial currents.
Matrix elements of the non-resonant vector current at $Q^2=0$
have been fixed through the previous analysis of
photon-induced meson-production data~\cite{knls13}.
We also need to fix the $Q^2$-dependence of the matrix elements
to study electron- and neutrino-induced reactions.
Regarding the axial current, 
we take advantage of the fact that most of our $\pi N\to MB$
($MB$: a meson-baryon state)
potentials are derived from a chiral Lagrangian.
Thus, we basically follow the way how the
axial current is introduced in the chiral Lagrangian:
an external axial current ($a^\mu_{\rm ext}$) enters into the chiral Lagrangian in
combination with the pion field as 
$\partial^\mu \pi +  f_\pi a^\mu_{\rm ext}$ where $f_\pi$ is the pion decay constant.
Then the tree-level non-resonant axial-current matrix elements are
derived from the chiral Lagrangian.
By construction, $A^{i,\mu}_{\rm NP,tree}$ ($i$: isovector component)
and the meson-baryon potential
$v$ satisfy the PCAC relation at $Q^2=-m^2_\pi$:
$\langle MB |q\cdot A^{i}_{\rm NP,tree} |  N\rangle = i f_\pi \langle MB | v | \pi^i N\rangle$.
The $Q^2$-dependence of the axial-coupling to the nucleon is fairly
well-known from data analyses of quasi-elastic neutrino scattering
and single pion electroproduction near threshold.
We employ the conventional dipole form factor,
$F_A(Q^2)=1/(1+Q^2/M^2_A)^2$, and take a numerical value for
the axial mass, $M_A=1.02$~GeV, from Ref.~\cite{axial-mass}.

\subsection{Matrix elements of $N^*$-excitation currents}

The hadronic vector current 
contributes to the neutrino-induced reactions
in the finite $Q^2$ region.
In Ref.~\cite{knls13},
we have done a combined analysis of 
$\pi N,\gamma p\to \pi N, \eta N, K\Lambda, K\Sigma$ reaction data, 
and fixed matrix elements of the vector current at $Q^2=0$ for the proton target. 
What we need to do is to extend the matrix elements of the vector current of Ref.~\cite{knls13}
to the finite $Q^2$ region for application to the neutrino reactions.
This can be done by analyzing 
data for electron-induced reactions on the proton and the neutron.
Then we separate the vector form factors for $N^*$ of $I=1/2$ ($I$: isospin)
into isovector and isoscalar parts.
Regarding $N^*$ of $I=3/2$ for which only the isovector current contributes,
we can determine the vector form factors by analyzing the proton-target data.

Because of rather scarce neutrino reaction data, it is difficult to
determine 
$N$-$N^*$ transition matrix elements induced by the axial-current.
This is in sharp contrast with the situation for the vector form factors
that are well determined by a large amount of electromagnetic reaction data.
Thus, we need to take a different path to fix the axial form factors. 
The conventional practice is 
to write down 
a $N$-$N^*$ transition
matrix element induced by the axial-current
in a general form with three or four form factors.
Then the PCAC relation,
$\langle N^*|q\cdot A^i_{\rm NP}|N\rangle
=if_\pi\langle N^*|\Gamma|\pi^i N\rangle$,
is invoked
to relate the presumably most important axial form factor at $Q^2=-m^2_\pi$
to the corresponding $\pi NN^*$ coupling.
The other form factors are ignored except for the pion pole term. 
We then assume 
$A^{i,\mu}_{\rm NP}(Q^2=-m^2_\pi)\sim A^{i,\mu}_{\rm NP}(Q^2=0)$.
In the present work, we consider 
the axial currents for bare $N^*$ of
the spin-parity $1/2^\pm$, $3/2^\pm$, $5/2^\pm$ and $7/2^\pm$,
and determine their axial form factors at $Q^2=0$ using the above procedure.
It is even more difficult to determine 
the $Q^2$-dependence of 
the axial couplings to $N$-$N^*$ transitions because of the limited
amount of data.
Thus we assume that
the $Q^2$-dependence of the axial form factors is the same as that used
for the non-resonant axial-current amplitudes.
It is worth emphasizing that a great advantage of our approach over
the existing models is that relative phases between resonant and
non-resonant amplitudes are made under control within the DCC model.
This is possible in our approach by constructing the axial-current amplitudes and
$\pi N$ interactions consistently with the requirement of the PCAC relation.

\section{Analysis of electron-induced reaction data}
\label{sec:electron}

Here we analyze data for electron-induced reactions off the proton
and neutron targets
to determine the $Q^2$ dependence of
the vector form factors.
The data we analyze span the kinematical region of 
$W\le$ 2~GeV and $Q^2\le$ 3~(GeV/$c$)$^2$
that is also shared by neutrino reactions for $E_\nu\le$ 2~GeV.

Among data for electron-proton reactions in the resonance region, 
those for the single pion electroproductions
are the most abundant
over a wide range of $W$ and $Q^2$.
Therefore, these
are the most useful to determine
the $Q^2$ dependence of the $p$-$N^*$ transition form factors.
The cross sections for $p(e,e'\pi^0)p$  and $p(e,e'\pi^+)n$ have
different sensitivities to resonances of different isospin state
($1/2$ or $3/2$).
The angular distribution of the pion is useful to disentangle the
spin-parity of the resonances.
Based on the one-photon exchange approximation,
a standard formula of the angular distribution for
the single pion electroproduction
can be expressed in terms of virtual photon cross sections
$\displaystyle {d\sigma_{\beta}(Q^2,W,\cos\theta_\pi^*)}/{d\Omega_\pi^*}$ ($\beta=T,L,LT,TT,LT'$).

The CLAS Collaboration has collected data
for the single pion electroproduction off the proton
in the kinematical region of our interest.
Then they have extracted from the data
the virtual photon cross sections.
We fit these virtual photon cross sections
to determine the $Q^2$ dependence of
the $p$-$N^*$ transition form factors.
The single pion electroproduction data occupy a substantial portion of the relevant
kinematical region of $W$ and $Q^2$.
In some kinematical region, however, 
we still need more data to fix the vector form factors.
In particular, data are missing for the $W\gtap 1.4$~GeV and low-$Q^2$ region,
and the $W\gtap 1.7$~GeV and $Q^2\ltap 2$ (GeV/$c$)$^2$.
In those kinematical region, 
we fit the inclusive structure functions from an empirical
model due to Christy and Bosted~\cite{christy}.

\begin{figure}[t]
\includegraphics[width=78mm]{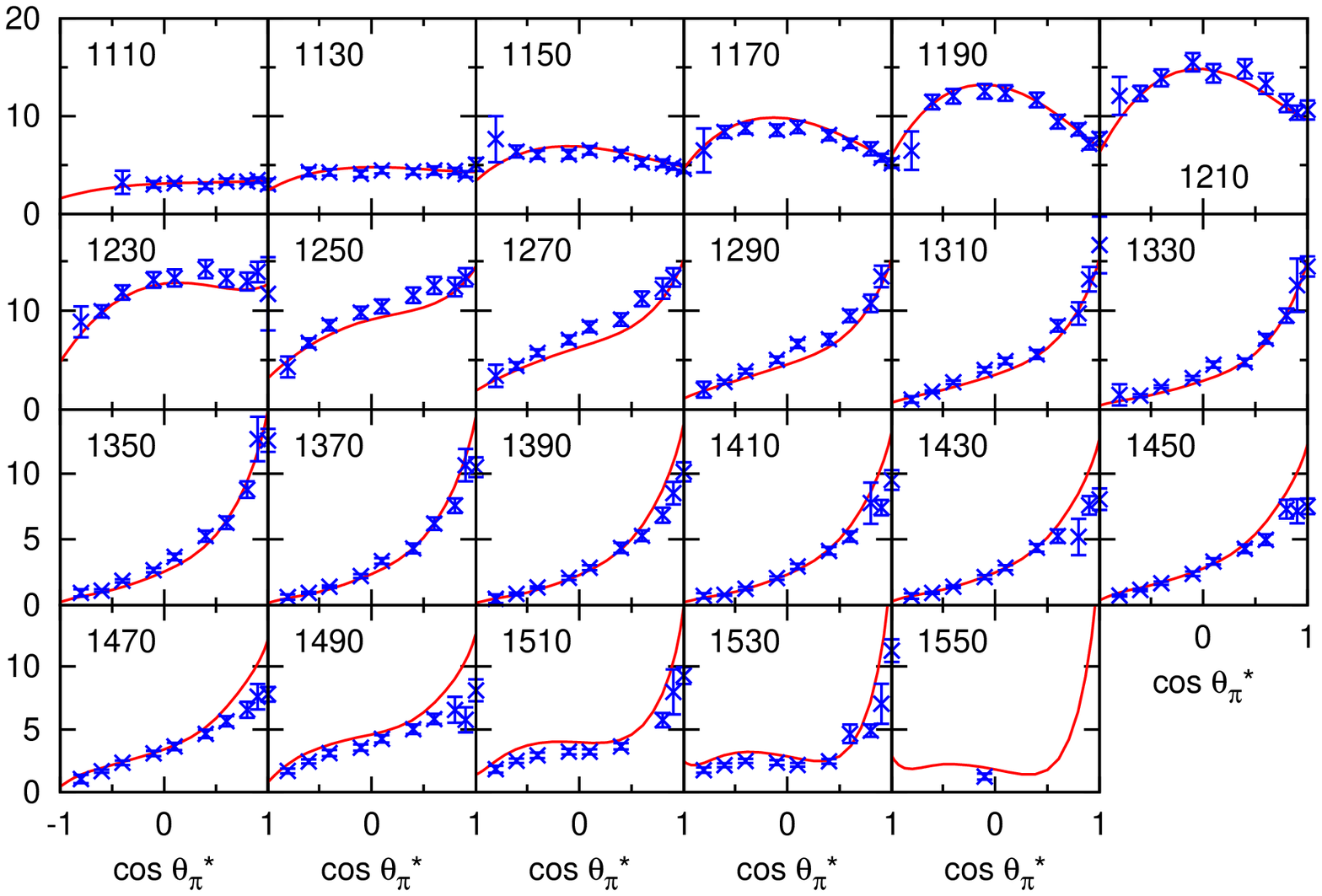}
\hspace{5mm}
\includegraphics[width=70mm]{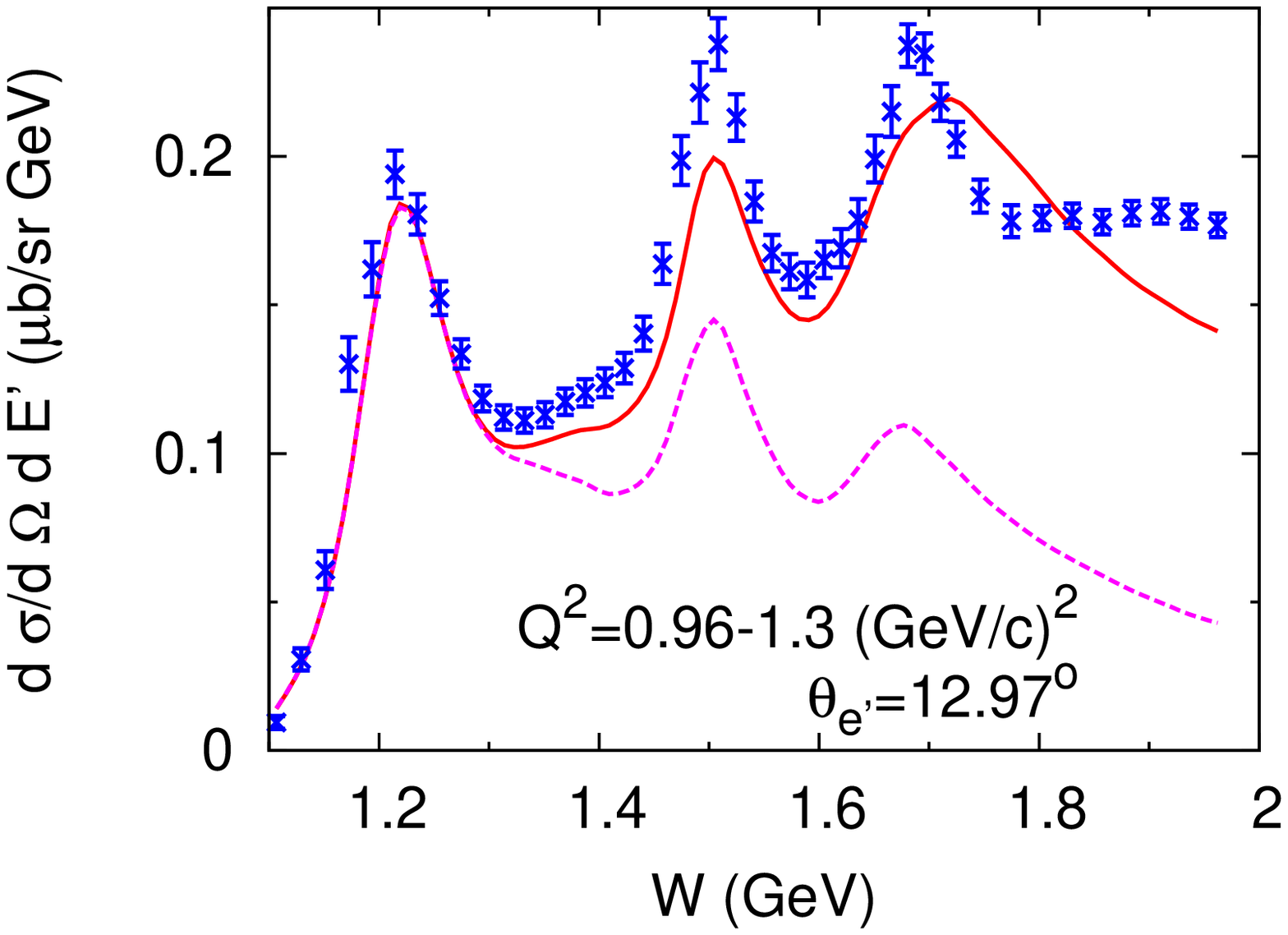}
\caption{(Color online)
(Left)
The virtual photon cross section 
$d\sigma_T/d\Omega^*_\pi+\epsilon\, d\sigma_L/d\Omega^*_\pi$ ($\mu$b/sr)
at $Q^2$=0.40 (GeV/$c$)$^2$ for
$p(e,e'\pi^+)n$ from the DCC model.
The number in each panel indicates $W$ (MeV).
The data are from Ref.~\cite{eepi-egiyan-prc}.
(Right)
Comparison of DCC-based calculation with data for inclusive electron-proton
 scattering at $E_e$=5.498~GeV.
The red solid curves are for inclusive cross sections while the magenta
dashed-curves 
are for contributions from the $\pi N$ final states.
The range of $Q^2$ and the electron scattering angle
 ($\theta_{e'}$) are indicated in each panel.
The data are from Ref.~\cite{E00-002}.
}
\label{fig:eepi-0.40}
\end{figure}

We have fitted the data at several $Q^2$ values where
the data are available. All the other parameters in the DCC model are
fixed as those determined in Ref.~\cite{knls13}.
We have successfully tested the DCC-based vector current model with the
data covering the 
whole kinematical region relevant to neutrino reactions of
$E_\nu\le 2$~GeV. 
We present a selected result for the analysis 
of electron-proton reactions.
We show a combination of the virtual photon cross sections,
$d\sigma_T/d\Omega^*_\pi+\epsilon\, d\sigma_L/d\Omega^*_\pi$,
at $Q^2$=0.40 (GeV/$c$)$^2$ 
for $p(e,e'\pi^+)n$ from the DCC model
in Fig.~\ref{fig:eepi-0.40} (left).
In the same figure, the corresponding data are also shown for
comparison. 
The DCC model fits the data for both $\pi^0$ and $\pi^+$
channels reasonably well. 
We also show in Fig.~\ref{fig:eepi-0.40} (right)
our DCC-based calculation of differential
cross sections of the inclusive electron-proton scattering 
in comparison with data;
the single pion electroproduction cross sections from the DCC
model are also presented.
In the figure, the range of $Q^2$ is indicated, and 
$Q^2$ monotonically decreases as $W$ increases.
The figures show a reasonable agreement between our calculation with the
data, and also show the increasing importance of the multi-pion production processes 
above the $\Delta(1232)$ resonance region.
As $Q^2$ increases, the DCC model starts to underestimate 
the inclusive cross section towards $W\sim$~2~GeV 
where the kinematical
region is entering the DIS 
and multi-meson production region.

Regarding the 
$\gamma n\to \pi N$ reactions,
we analyze unpolarized differential cross sections data
from $\pi N$ threshold to $W=2$~GeV,
and determine the vector $nN^*$ transition strengths at
$Q^2$=0 for $I$=1/2 $N^*$ states.
In the finite $Q^2$ region,
we use empirical inclusive structure functions from Ref.~\cite{bosted,christy2} as data
to determine the transition vector form factors.
We successfully fitted the data by adjusting the vector form factors.
See Ref.~\cite{dcc-neutrino} for numerical results.

\section{Results for neutrino reactions}
\label{sec:results}

\begin{figure}[t]
\includegraphics[height=0.33\textwidth]{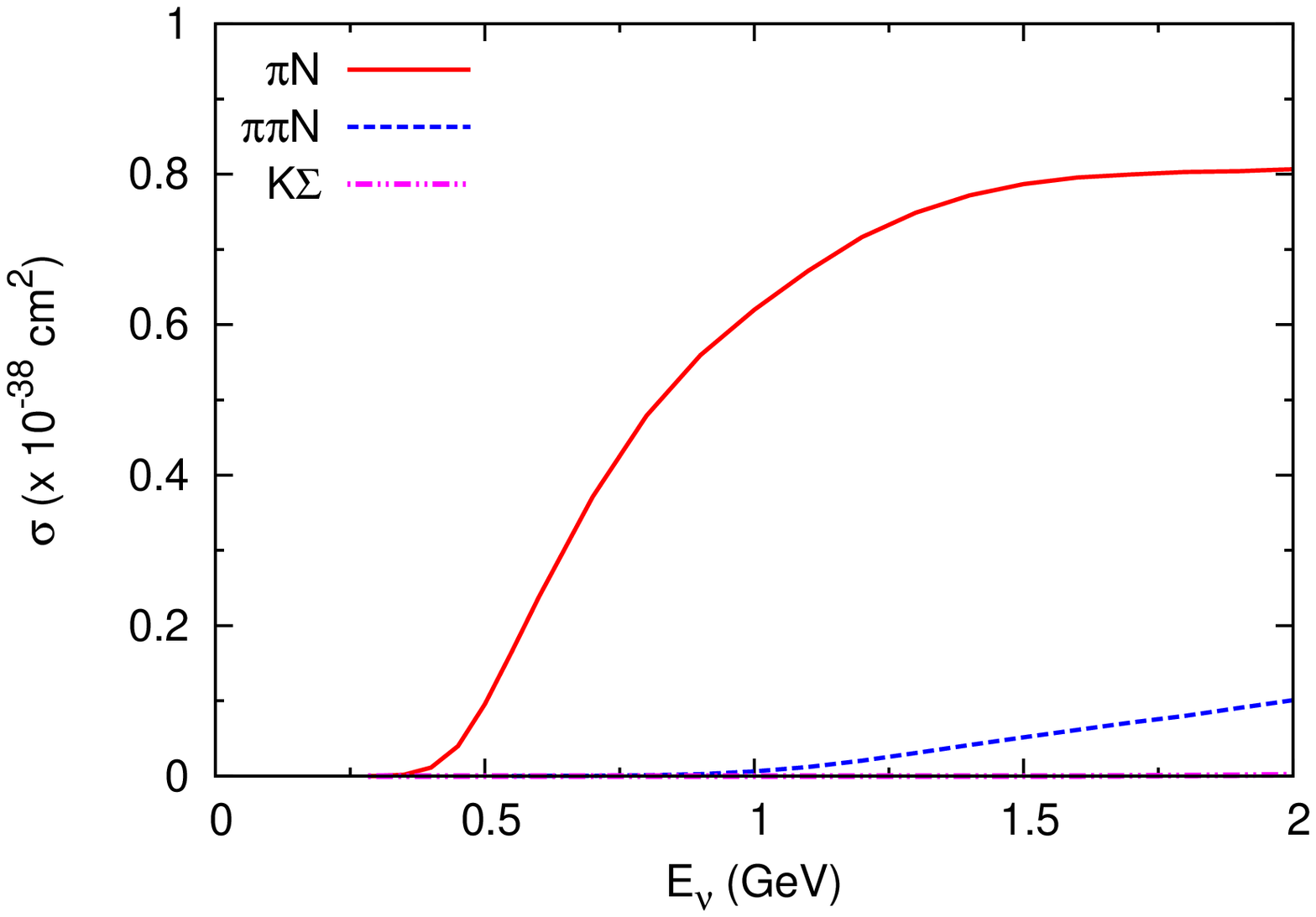}
\includegraphics[height=0.33\textwidth]{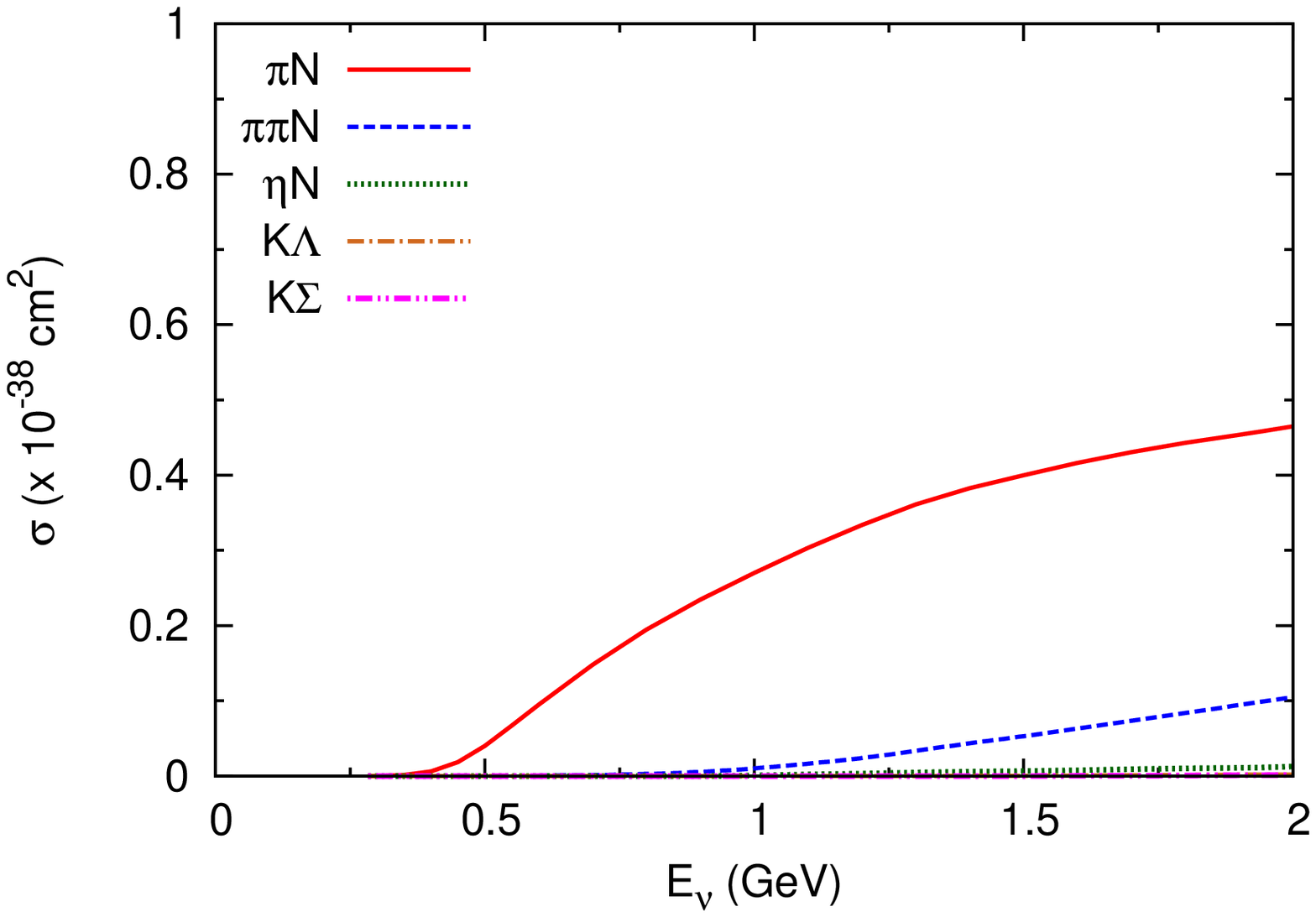}
\caption{(Color online)
Total cross sections for the CC $\nu_\mu\, p$ (left)
and $\nu_\mu n$ (right) reactions.
}
\label{fig:neutrino-tot}
\end{figure}
We present cross sections for the
$\nu_\mu\, N$ reactions.
With the DCC model, we can predict contributions from all the final states
included in our model.
Also, the DCC model provides all possible differential cross sections
for each channel.
Here, we present total cross sections for the CC
$\nu_\mu\, N$ reactions up to $E_\nu=2$~GeV
in Fig.~\ref{fig:neutrino-tot}.
For the proton-target, the single pion production dominates in the
considered energy region.
For the neutron-target, the single pion production is still the largest,
but double-pion production becomes relatively more important towards $E_\nu=2$~GeV.
The $\eta N$ and $KY$ production cross sections are ${\cal O}(10^{-1}$-$10^{-2})$ smaller.

\begin{figure}[t]
\includegraphics[height=0.33\textwidth]{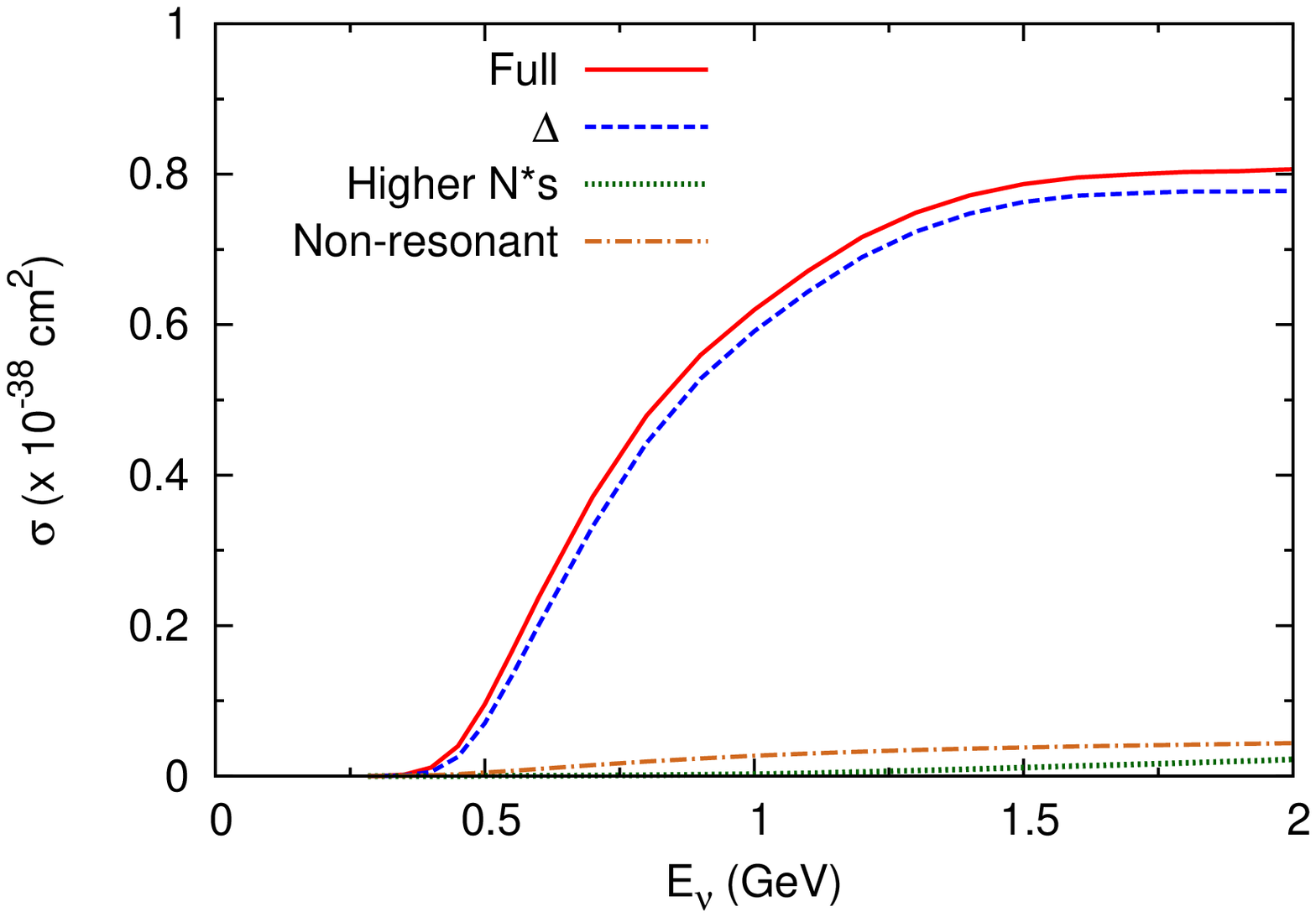}
\includegraphics[height=0.33\textwidth]{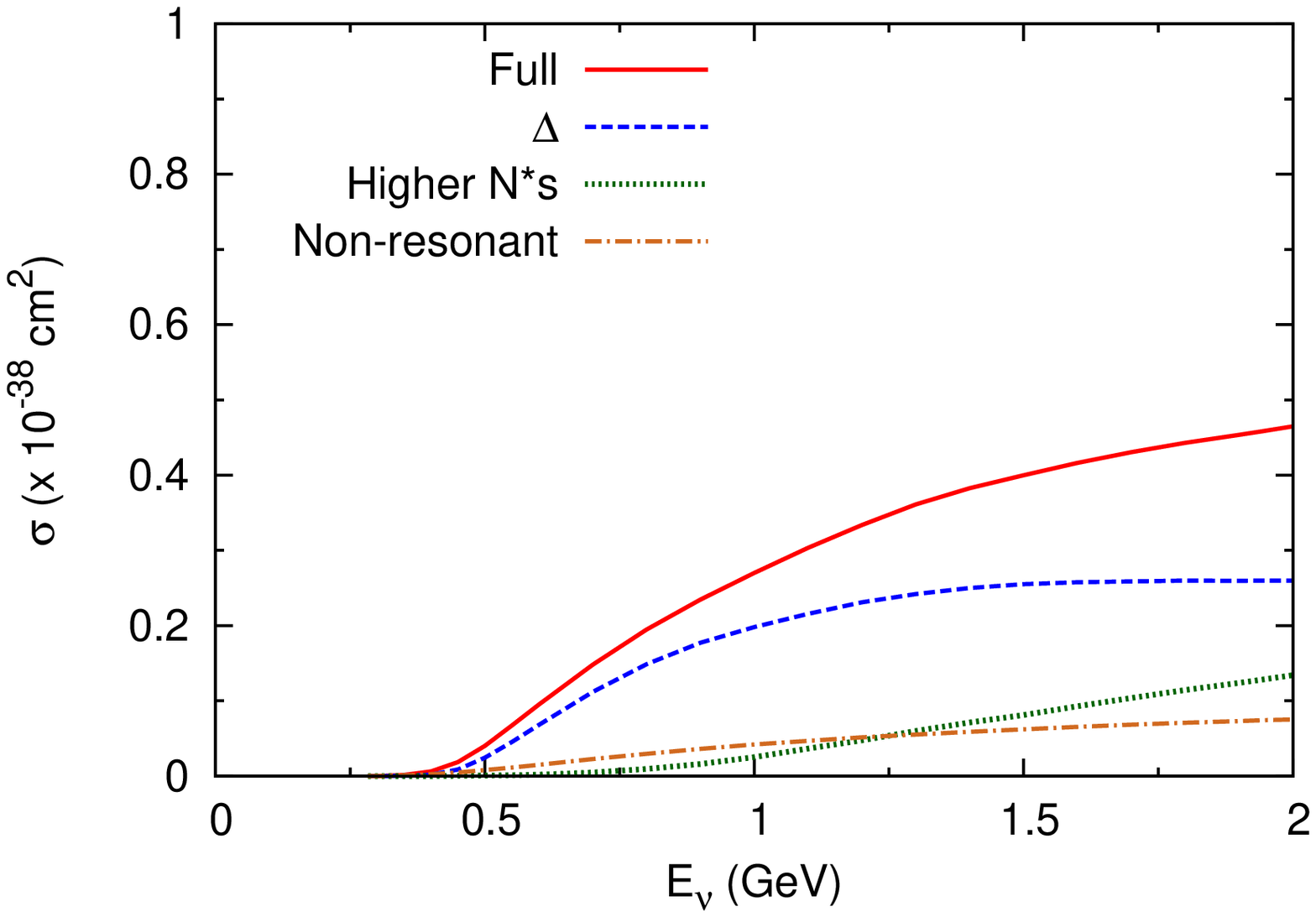}
\caption{(Color online)
Various mechanisms contributing to
$\nu_\mu\, p\to \mu^- \pi^+ p$ (left)
and $\nu_\mu n\to \mu^- \pi N$ (right).
}
\label{fig:neutrino-tot-comp}
\end{figure}
Next we examine reaction mechanisms of the $\nu_\mu\, N$ scattering.
In Fig.~\ref{fig:neutrino-tot-comp}, we break down the single-pion
production cross sections into several contributions each of which contains a
set of certain mechanisms.
For the proton-target process, 
the contribution from the $\Delta$(1232) resonance dominates,
while the higher $N^*$ contribution is very small.
The $\Delta$ contribution here is the neutrino cross section calculated
with the $P_{33}$ partial wave amplitude that
contains the $N^*$-excitation mechanisms,
while the higher $N^*$ contribution is from 
the resonant amplitude including all partial waves other than
$P_{33}$.
The non-resonant cross sections
calculated from the non-resonant amplitude
is small for the proton-target process.
In contrast, the situation is more complex in the neutron-target
process where the $\Delta$ gives a smaller contribution and 
both $I=$1/2 and 3/2 resonances contribute.
As can be seen in the right panel of 
Fig.~\ref{fig:neutrino-tot-comp},
the $\Delta$ dominates for $E_\nu\ltap 1$~GeV, 
and higher resonances and non-resonant mechanisms give comparable
contributions towards $E_\nu\sim 2$~GeV.
This shows an importance of including both resonant and non-resonant
contributions with the interferences among them under control.
Similarly, 
we can compare
the contribution of resonant and non-resonant amplitudes
for the two-pion production reaction.
Because $\Delta(1232)$ mainly contributes below the $\pi\pi N$
production threshold and thus gives a small contribution here,
the resonant and non-resonant contributions are more comparable.
Still, we find that the resonance-excitations are the main
mechanism for the double-pion production in the resonance region.

\begin{figure}[t]
\includegraphics[height=0.24\textwidth]{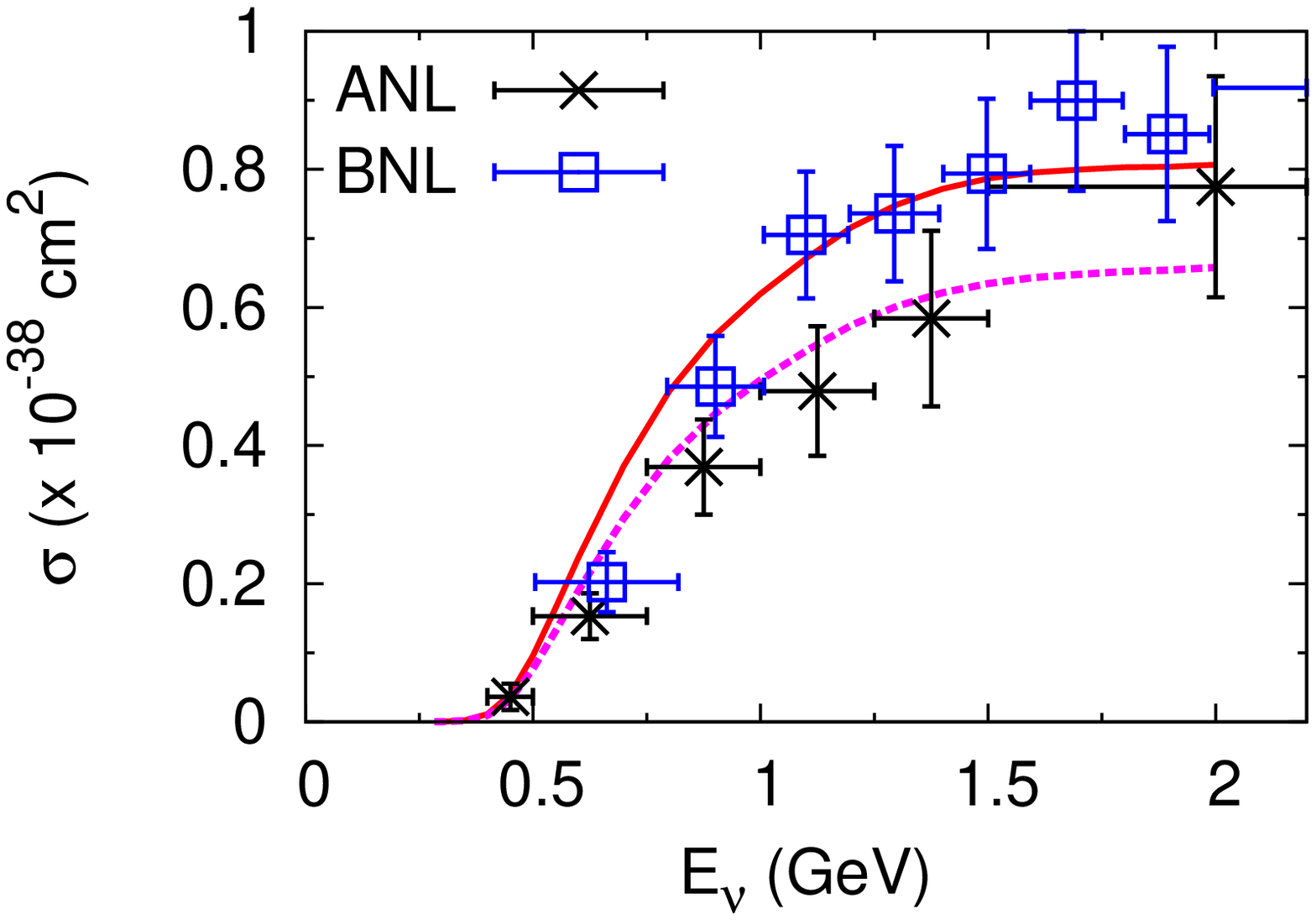}
\hspace{-5mm}
\includegraphics[height=0.24\textwidth]{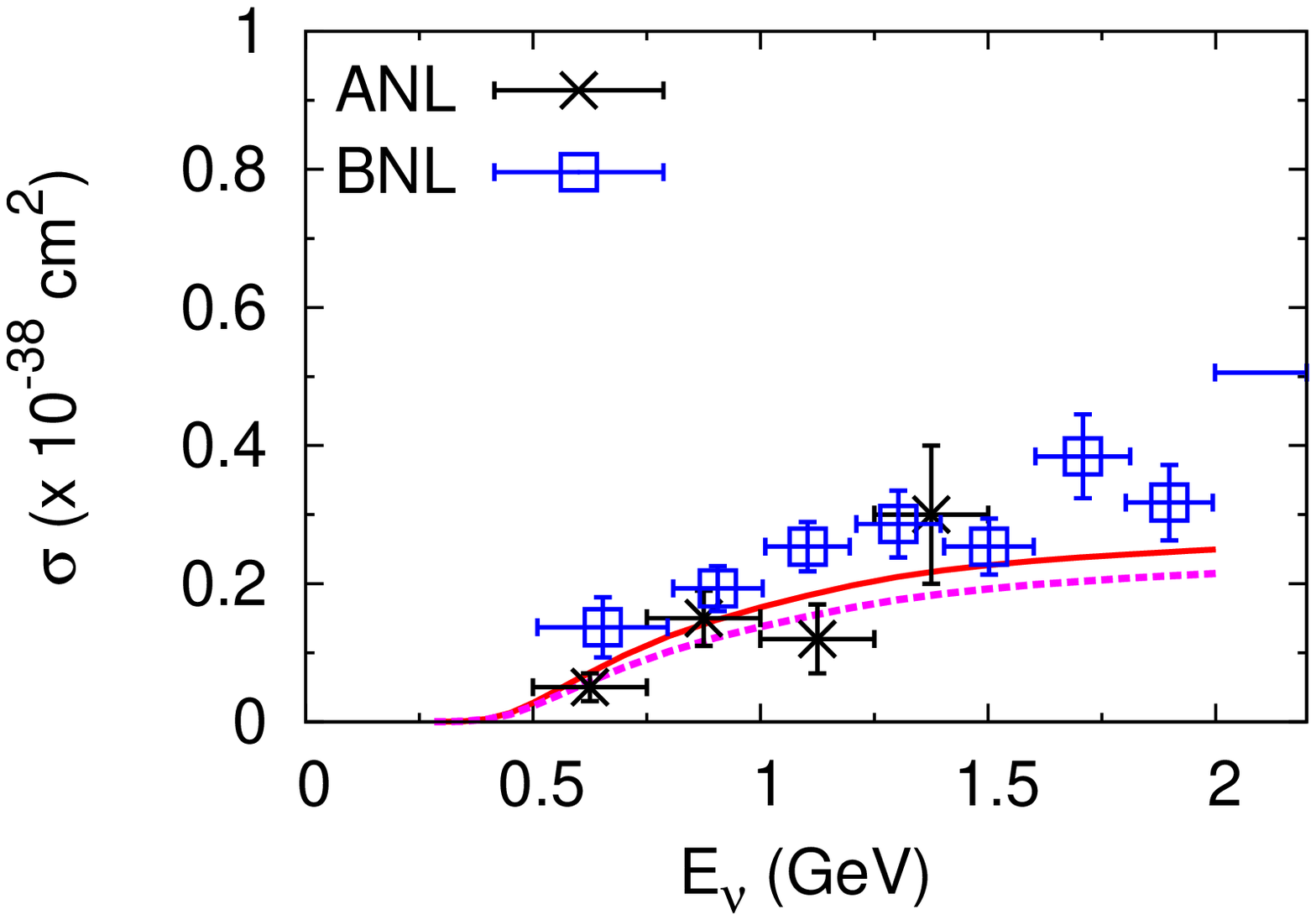}
\hspace{-5mm}
\includegraphics[height=0.24\textwidth]{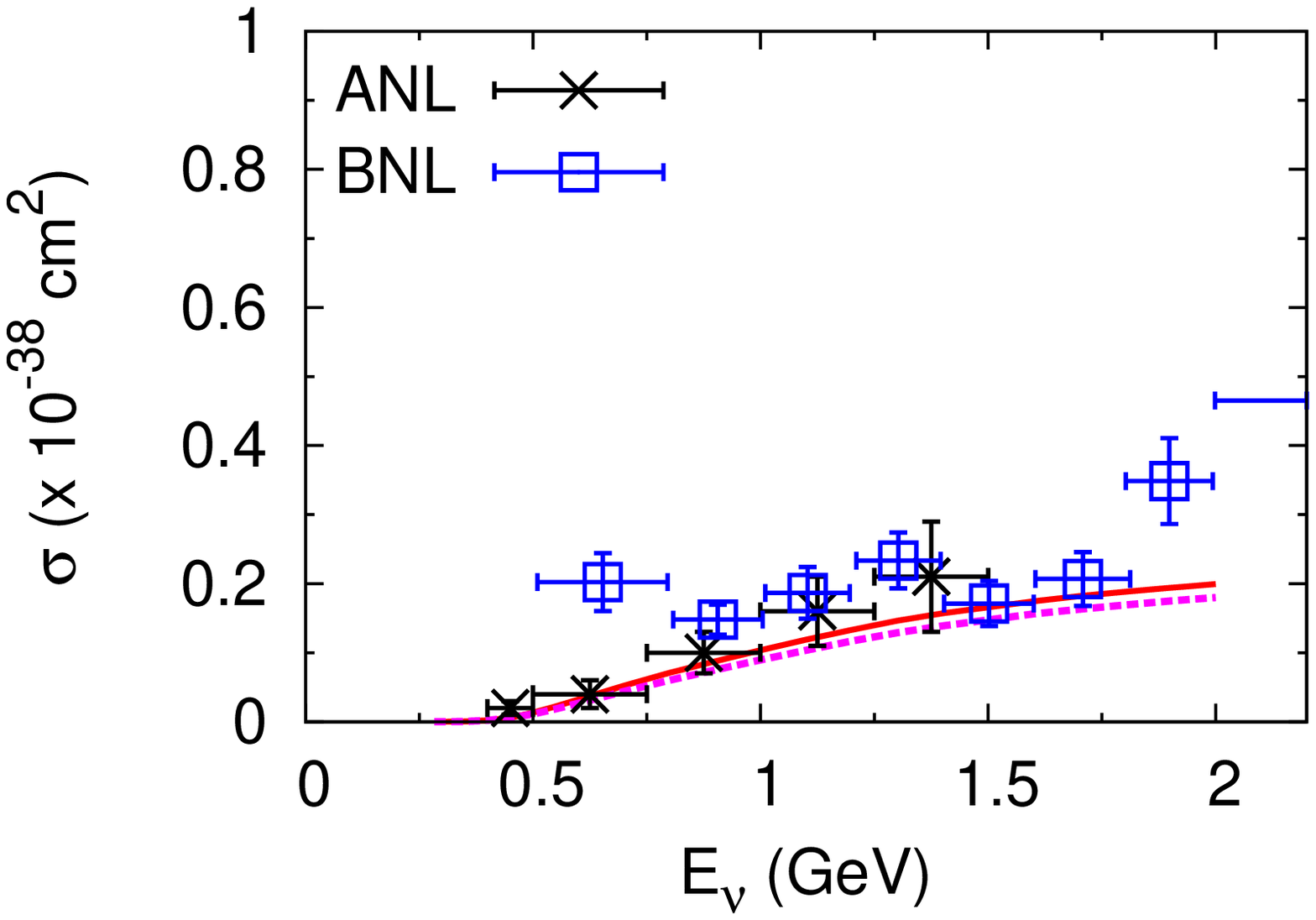}
\caption{(Color online)
Comparison of the DCC-based calculation (red solid curves)
with data for 
$\nu_\mu\, p\to \mu^- \pi^+ p$ (left),
$\nu_\mu n\to \mu^- \pi^0 p$ (middle)
and $\nu_\mu n\to \mu^- \pi^+ n$ (right).
The DCC calculation with $0.8\times g_{AN\Delta(1232)}^{\rm PCAC}$
is also shown (magenta dashed curve).
ANL (BNL) data are from Ref.~\cite{anl} (\cite{bnl}).
}
\label{fig:neutrino-tot-data}
\end{figure}
Next we compare the CC neutrino-induced single pion production cross
sections from the DCC model with
available data from Refs.~\cite{anl,bnl} in Fig.~\ref{fig:neutrino-tot-data}.
The left panel shows the total cross sections for
$\nu_\mu\, p\to \mu^- \pi^+ p$ for which $\Delta(1232)$ dominates as we
have seen in Fig.~\ref{fig:neutrino-tot-comp}.
If the $\Delta(1232)$-dominance persists in the neutron-target processes
shown in the middle and right panels of
Fig.~\ref{fig:neutrino-tot-data},
the isospin Clebsch-Gordan coefficients determine the relative strength
as
$\sigma(\nu_\mu n\to \mu^- \pi^0 p)/\sigma(\nu_\mu p\to \mu^- \pi^+ p) = 2/9 \sim 0.22$, and
$\sigma(\nu_\mu n\to \mu^- \pi^+ n)/\sigma(\nu_\mu p\to \mu^- \pi^+ p) = 1/9 \sim 0.11$.
The actual ratios from the DCC model are
$\sigma(\nu_\mu n\to \mu^- \pi^0 p)/\sigma(\nu_\mu p\to \mu^- \pi^+ p) =$
0.28, 0.27, 0.29, and 
$\sigma(\nu_\mu n\to \mu^- \pi^+ n)/\sigma(\nu_\mu p\to \mu^- \pi^+ p) =$
0.13, 0.17, 0.21
at $E_\nu$=0.5, 1, 1.5~GeV, respectively.
The deviations from the naive isospin analysis are due to the
the non-resonant and higher-resonances 
contributions mostly in the neutron-target processes,
as we have seen in Fig.~\ref{fig:neutrino-tot-comp}.
The two datasets from BNL and ANL 
for $\nu_\mu p\to \mu^- \pi^+ p$
shown in the left panel of
Fig.~\ref{fig:neutrino-tot-data} are not consistent
as has been well known,
and our result is closer to the BNL data~\cite{anl}.
For the other channels, our result is fairly consistent with both of the
BNL and ANL data.
It seems that 
the bare axial $N$-$\Delta(1232)$ coupling constants
determined by the PCAC relation are too large to 
reproduce the ANL data.
Because 
axial $N$-$N^*$ coupling constants should be better determined by analyzing
neutrino-reaction data,
it is tempting to 
multiply the bare axial
$N$-$\Delta(1232)$ coupling constants,
$g_{AN\Delta(1232)}^{\rm PCAC}$,
by 0.8, so that the DCC model better fits the ANL data.
The resulting cross sections are shown by the dashed curves in 
Fig.~\ref{fig:neutrino-tot-data}.
We find that
$\sigma(\nu_\mu p\to \mu^- \pi^+ p)$ is reduced due to the
dominance of the $\Delta(1232)$ resonance in this channel, while 
$\sigma(\nu_\mu n\to \mu^- \pi N)$ is only slightly reduced.
The original data of these two experimental data have been
reanalyzed recently~\cite{reanalysis}, and it was
claimed that the discrepancy between the two datasets is resolved. 
The resulting cross sections are closer to the original ANL data.
However, the number of data is still very limited, and 
a new measurement of neutrino cross sections on the hydrogen and
deuterium is highly desirable.
We also note that the data shown in Fig.~\ref{fig:neutrino-tot-data} were taken from
experiments using the deuterium target.
Thus one should analyze the data considering the nuclear effects such as
the initial two-nucleon correlation and the final state interactions.
Recently, the authors of Ref.~\cite{wsl} have taken a first step towards 
such an analysis.
They developed a model that consists of
elementary amplitudes for neutrino-induced single pion production off the
nucleon~\cite{sul}, pion-nucleon rescattering amplitudes, and 
the deuteron and final $NN$ scattering wave functions.
Although they did not analyze the ANL and BNL data with their model, 
they examined how much the cross sections at certain kinematics
can be changed by considering the nuclear effects. 
They found that the cross sections can be reduced as much as 30\% 
for $\nu_\mu d\to \mu^- \pi^+ pn$ due to the $NN$ rescattering.
Meanwhile, the cross sections 
for $\nu_\mu d\to \mu^- \pi^0 pp$ are hardly changed by the final state
interaction. 
It will be important to analyze the ANL and BNL data with this kind of model to
determine the axial nucleon current, particularly the axial
$N$-$\Delta$(1232) transition strength.

\begin{figure}[t]
\includegraphics[height=0.24\textwidth]{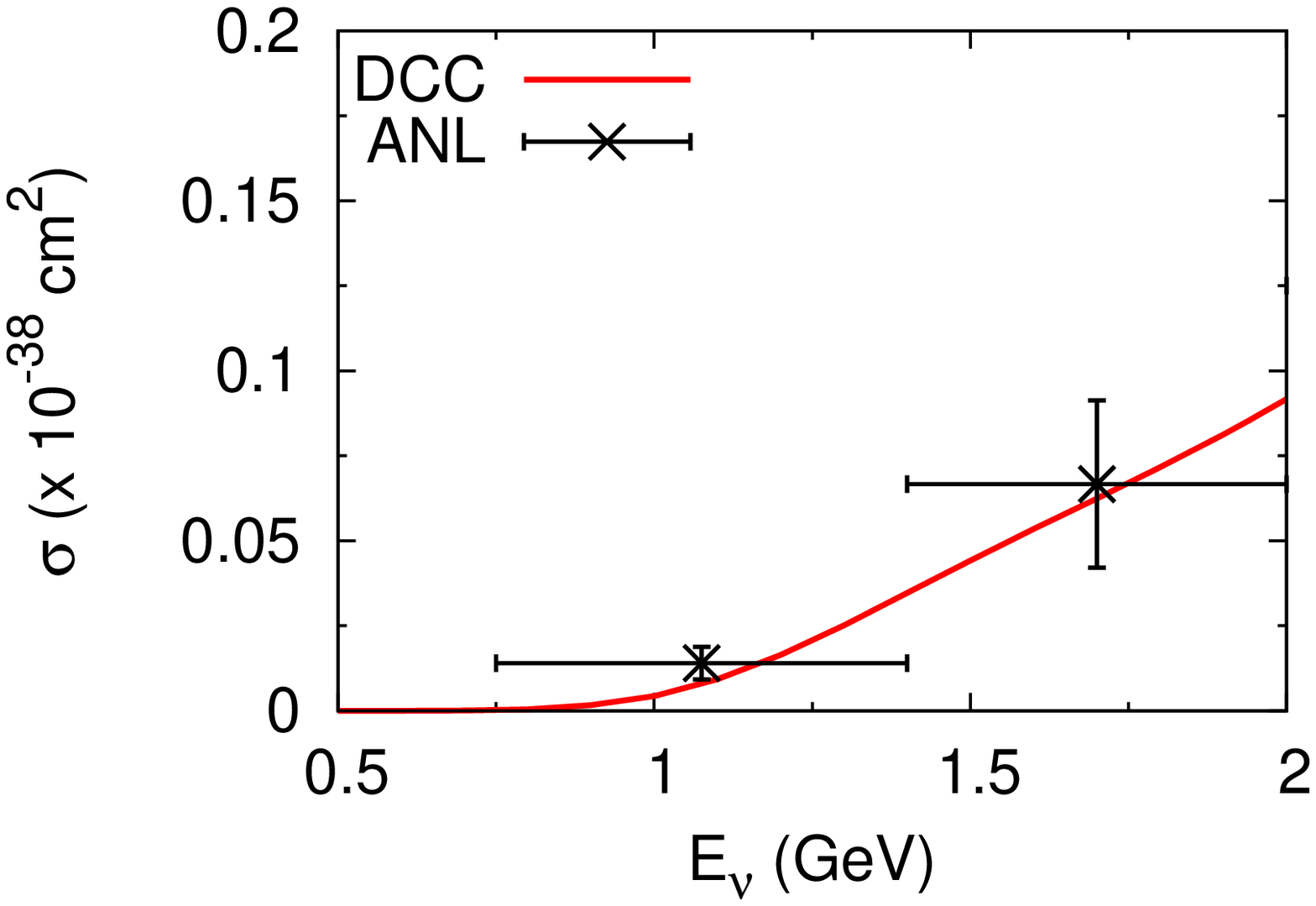}
\hspace{-5mm}
\includegraphics[height=0.24\textwidth]{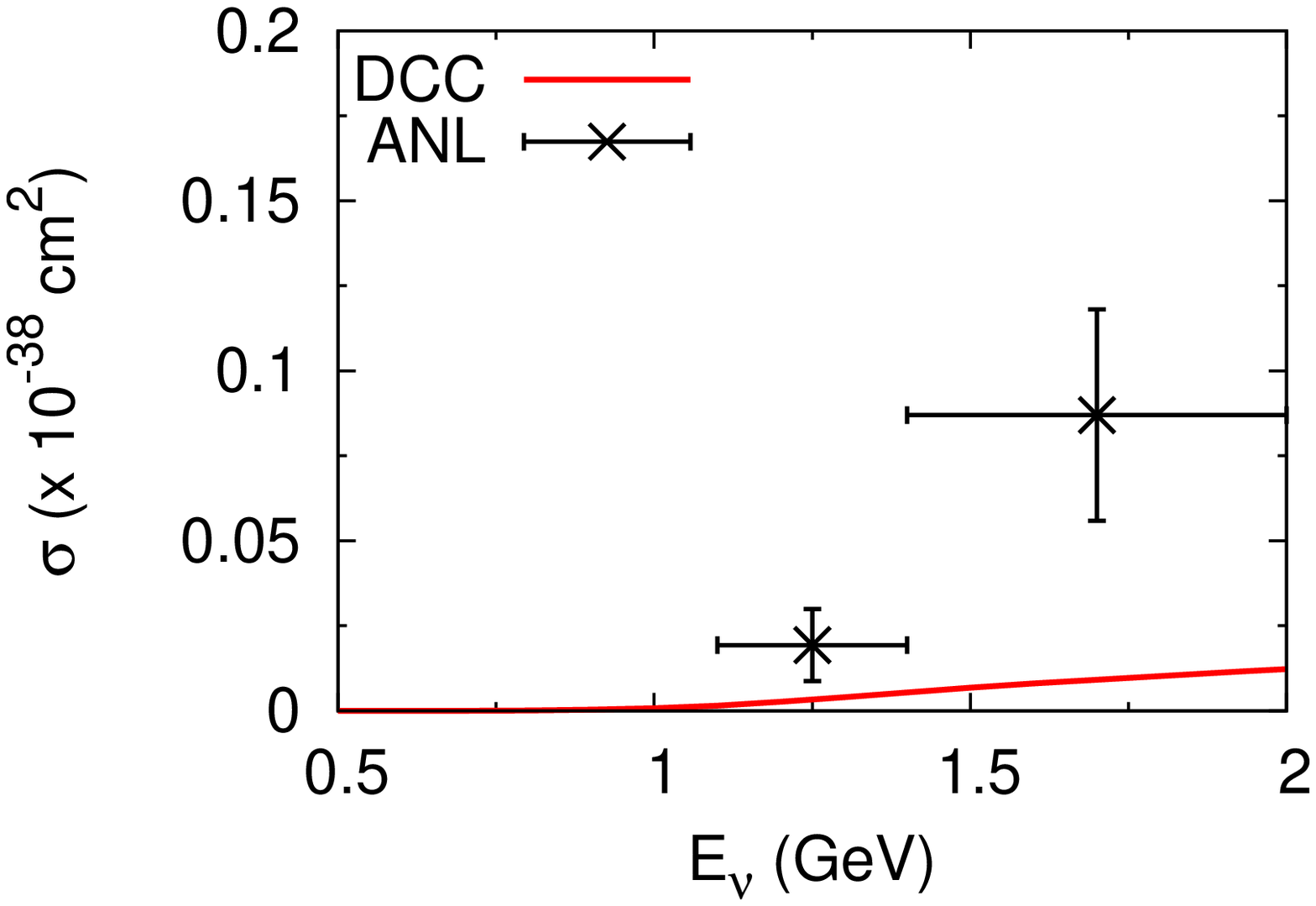}
\hspace{-5mm}
\includegraphics[height=0.24\textwidth]{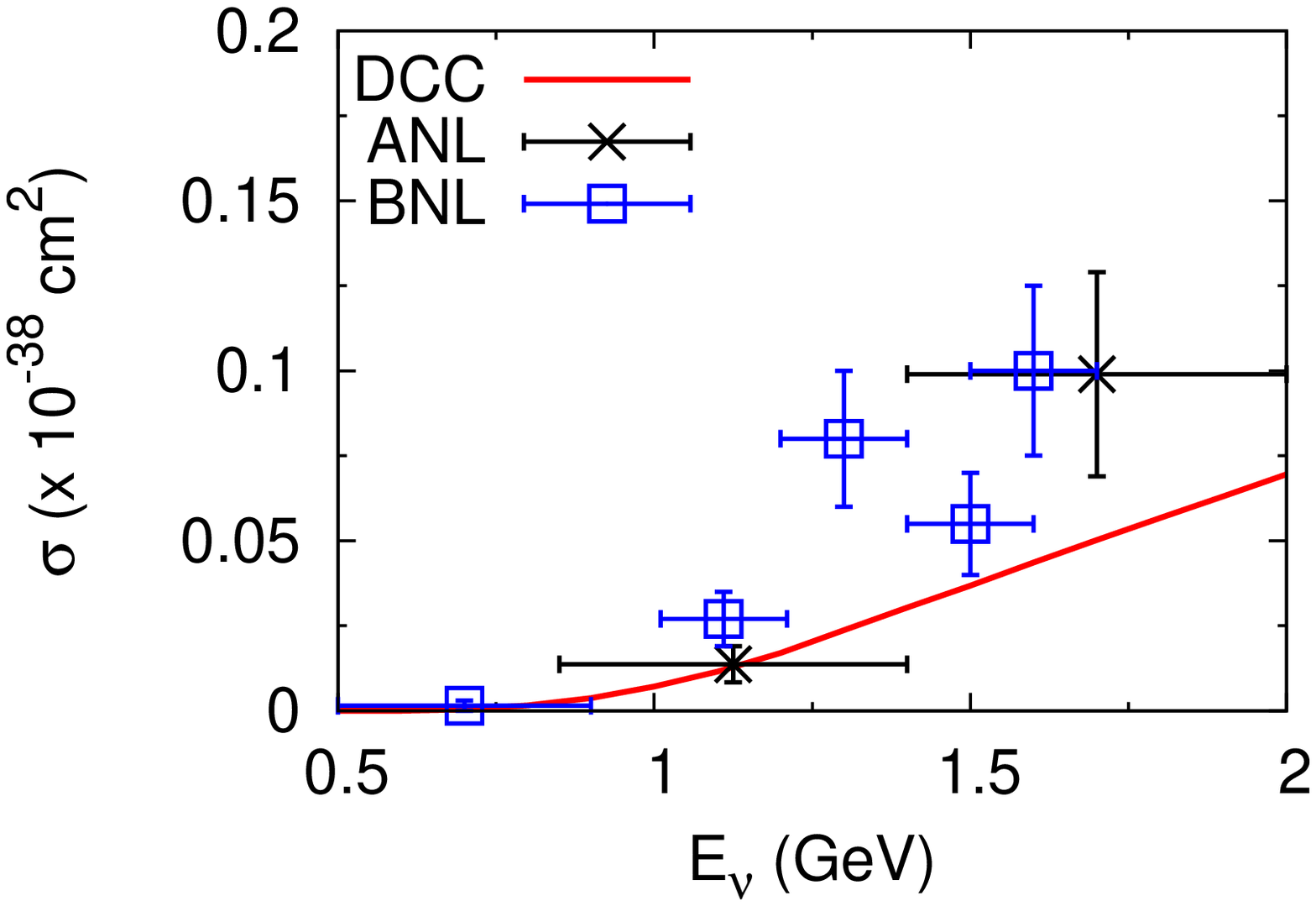}
\caption{(Color online)
Comparison of the DCC-based calculation with data for 
$\nu_\mu\, p\to \mu^- \pi^+\pi^0 p$ (left),
$\nu_\mu p\to \mu^- \pi^+\pi^+ n$ (middle)
and $\nu_\mu n\to \mu^- \pi^+\pi^- p$ (right).
ANL (BNL) data are from Ref.~\cite{anl2} (\cite{bnl}).
}
\label{fig:neutrino-pipin-data}
\end{figure}
We finally compare our results for double-pion productions with existing
data in Fig.~\ref{fig:neutrino-pipin-data}.
Although there exist a few theoretical works on the neutrino-induced double-pion
production near threshold~\cite{biswas,adjei,spain-pipin},
our calculation for the first time takes account of relevant resonance
contributions for this process.
The DCC-based prediction is fairly consistent with the data in the order
of the magnitude.
Particularly,
the cross sections for $\nu_\mu\, p\to \mu^- \pi^+\pi^0 p$ from the DCC
model are in agreement with data.
However, the DCC prediction underestimates 
the $\nu_\mu\, p\to \mu^- \pi^+\pi^+ n$ data.
The rather small ratio of
$\sigma(\nu_\mu\, p\to \mu^- \pi^+\pi^+ n) /
\sigma (\nu_\mu\, p\to \mu^- \pi^+\pi^0 p)\sim 13\%$
at $E_\nu$=2~GeV from our calculation
can be understood as follows.
Within the present DCC-based calculation, $\pi\pi N$ final states are 
from decays of 
the $\pi N$ and
of the $\pi\Delta$, $\rho N$, $\sigma N$ quasi two-body states.
For a neutrino CC process on the proton for which hadronic states have
$I=3/2$, the $\pi N$, $\pi\Delta$, $\rho N$ channels can contribute.
Within the current DCC model, we found that the $\pi\Delta$ channel
gives a dominant contribution to the double pion productions.
Then, retaining only the $\pi\Delta$ contribution,
the ratio is given by
the isospin Clebsch-Gordan coefficients as,
$\sigma(\nu_\mu\, p\to \mu^- \pi^+\pi^+ n) /
\sigma (\nu_\mu\, p\to \mu^- \pi^+\pi^0 p)=2/13\sim 15\%$,
in good agreement with the ratio from the full calculation.
With a very limited dataset, we do not further pursue the origin
of the difference between our calculation and the data.
If the double-pion data are further confirmed, 
then the model needs to incorporate some other mechanisms
and/or adjust model parameters of the DCC model to explain the data.

\section{Summary}

In this work, we have developed a dynamical coupled-channels (DCC) model for
neutrino-nucleon reactions in the resonance region.
Our starting point is the DCC model that we have developed through a
comprehensive analysis of 
$\pi N, \gamma p\to \pi N, \eta N, K\Lambda, K\Sigma$ data
for $W\le 2.1$~GeV~\cite{knls13}.
In order to extend the DCC model of Ref.~\cite{knls13} to what works for the neutrino
reactions, we analyzed data for
the single pion photoproduction off the neutron, 
and also data for the electron scattering on both
proton and neutron targets.
Through the analysis,
we determined the $Q^2$-dependence of the
vector form factors up to $Q^2\le 3$~(GeV/$c$)$^2$. 
We derive the axial-current matrix elements 
that are linked to the $\pi N$ potentials of the DCC model
through the PCAC relation.
As a consequence, relative phases between the non-resonant and resonant
axial current amplitudes are uniquely determined within the DCC model.

We have presented cross sections for the neutrino-induced
meson productions for $E_\nu\le 2$~GeV.
In this energy region, the single-pion production gives the largest
contribution.
Towards $E_\nu\sim 2$~GeV, the cross section for the double-pion production
is getting larger to become 1/8 (1/4) of the single-pion production
cross section for the proton (neutron) target.
Because our DCC model has been determined by analyzing 
the $\pi N, \gamma N\to \pi N, \eta N, K\Lambda, K\Sigma$ data,
we can also make a quantitative prediction for the neutrino cross sections
for $\eta N$, $K\Lambda$, and $K\Sigma$ productions.
We found that
cross sections for $\eta N, K\Lambda$ and $K\Sigma$ productions are
$10^{-2}$-$10^{-3}$ times smaller than those for the single pion production.
We have compared our numerical results with the available experimental
data. 
For the single-pion production, our result,
for which the axial $N$-$N^*$ couplings are fixed by the PCAC relation,
is consistent with the BNL
data for $\nu_\mu p\to\mu^-\pi^+p$,
while fair agreement with both ANL
and BNL data is found for the neutron target data.
Through the comparison with the single pion production data for 
$W\ltap$1.4~GeV for which 
the $\Delta(1232)$-excitation is the dominant mechanism,
we were able to study the strength and the $Q^2$-dependence of the axial
$N$-$\Delta(1232)$ coupling.
We also calculated double-pion production cross sections by taking
account of relevant resonance contributions for the first time, and
compared them with the data.
We found a good agreement for
$\nu_\mu\, p\to \mu^- \pi^+\pi^0 p$ and 
$\nu_\mu n\to \mu^- \pi^+\pi^- p$, but not for 
$\nu_\mu p\to \mu^- \pi^+\pi^+ n$.
Because the data for the double-pion productions are statistically rather poor,
it is difficult to make a conclusive judgement on the DCC model.

\begin{center}
{\large\bf Acknowledgments}
\end{center}
The author thanks Hiroyuki Kamano and Toru Sato for their collaboration.
This work was
supported by Ministry of Education, Culture, Sports, Science and
Technology (MEXT) KAKENHI Grant Number 25105010.

\end{document}